\documentstyle[psfig]{res}
\begin{document}

\title{PRIMORDIAL\, DEUTERIUM\\  ABUNDANCE\, MEASUREMENTS}

\author{Sergei A. LEVSHAKOV \\
{\it National Astronomical Observatory, Mitaka, 
Tokyo 181-8558,  JAPAN, lev@diamond.mtk.nao.ac.jp}\\
Wilhelm H. KEGEL\\
{\it Institut f\"ur Theoretische Physik der Universit\"at
Frankfurt am Main, Postfach 11 19 32, 60054 Frankfurt/Main 11,
GERMANY, kegel@astro.uni-frankfurt.de}\\
Fumio TAKAHARA\\
{\it Department of Earth and Space Science, Faculty of Science,
Osaka University, Toyonaka, Osaka 560, JAPAN, 
takahara@vega.ess.sci.osaka-u.ac.jp}}

\maketitle

\section*{Abstract}

Deuterium abundances measured recently from QSO absorption-line systems
lie in the range from $\simeq 3\times10^{-5}$ to 
$\simeq 3\times10^{-4}$, which shed some questions on
standard big bang theory.
We show that this discordance may simply be an
{\it artifact} caused by inadequate analysis ignoring
spatial correlations in the velocity field in turbulent media.
The generalized procedure (accounting for such correlations)
is suggested to reconcile the D/H measurements.

An example is presented based on two high-resolution observations
of Q~1009+2956 (low D/H) [1,2] and Q~1718+4807 (high D/H) [8,9]. 
We show that both observations are compatible with
D/H $\simeq 4.1 - 4.6\times10^{-5}$, and thus support SBBN.
The estimated mean value $\langle$ D/H $\rangle \simeq 4.4\times10^{-5}$
corresponds to the baryon-to-photon ratio during SBBN
$\eta \simeq 4.4\times10^{-10}$ which yields the present-day baryon
density $\Omega_b h^2 \simeq 0.015$.

\section{Introduction}

From recent HST observations of the 
$z_a = 0.7$ absorption-line system toward the quasar Q~1718+4807
Webb {\it et al.} [8,9] deduced D/H = $1.8-3.1\times10^{-4}$. 
This ratio is significantly higher than that derived from other 
quasar spectra. For instance, at $z_a = 2.504$ 
toward Q~1009+2956 the D/H ratio lies in the range from 
$1.8\times10^{-5}$ to $3.5\,10^{-5}$ [1]. 
New measurements [2] give a slightly higher range for the D/H value
at $z_a = 2.504$ ($3.31 - 4.57\times10^{-5}$) which is still 
incompatible with the results of Webb {\it et al.}

\begin{figure*}
\vspace{0.0cm}
\hspace{0.0cm}\psfig{figure=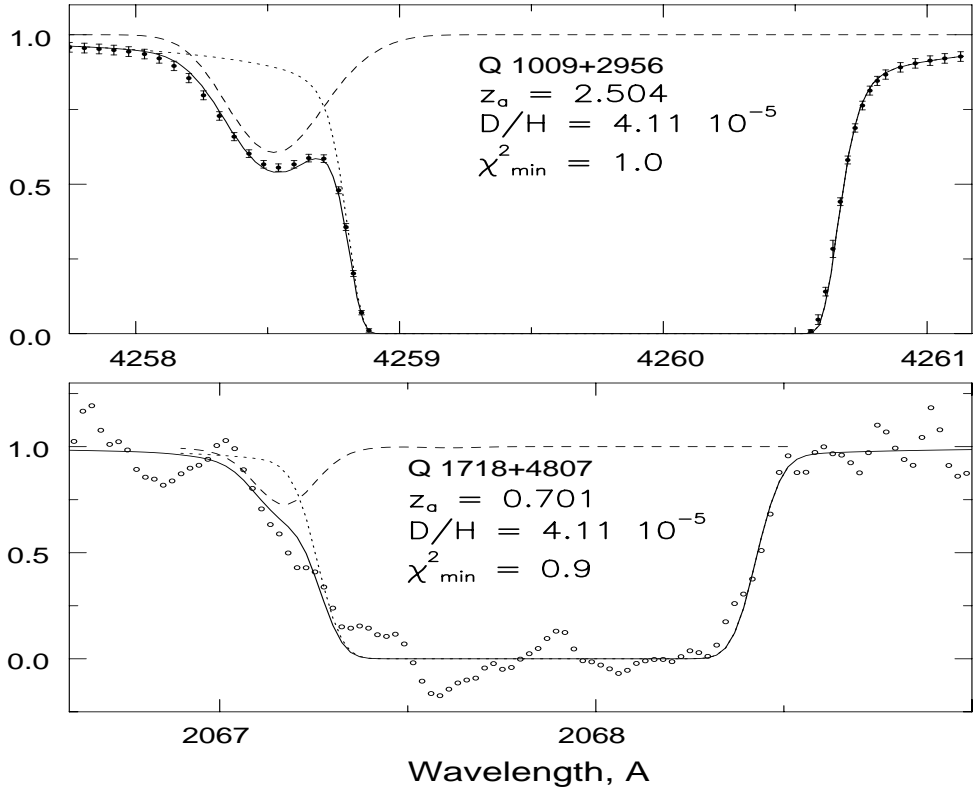,height=7cm,width=12.0cm}
\vspace{1.0cm}
\caption[]{{\it Upper panel.} An example of a template H+D Ly$\alpha$
profile (dots) at $z_a = 2.504$ toward the quasar Q~1009+2956
representing the normalized intensities and their uncertainties
(error bars) in accord with [1]. The results of the RMC minimization
are shown by the solid curve (H+D), the dotted curve (H~I) and the dashed curve 
(D~I)
for the mesoturbulent model with N(H~I) $= 2.71\times10^{17}$ cm$^{-2}$,
D/H = $4.11\times10^{-5}$, $T_{kin} = 20220$ K, $\sigma_t = 25.8$
km s$^{-1}$, and $L/l = 4.1$ .\\
{\it Lower panel.} HST/GHRS observations [8,9] of the H+D Ly$\alpha$ line
at $z_a = 0.701$ toward the quasar Q~1718+4807 
(the open circles represent
the normalized flux, the spectral resolution is 0.1 \AA ). Profiles for
H~I (dotted curve), D~I (dashed curve) and H+D (solid curve) 
are computed
by the RMC procedure for N(H~I) = $1.77\times10^{17}$ cm$^{-2}$,
D/H = $4.11\times10^{-5}$, $T_{kin} = 15100$ K, $\sigma_t = 26$
km s$^{-1}$, and $L/l = 3.5$ .
The $\chi^2$ values (per degree of freedom) of the final fits are
shown. 
}
\end{figure*}

According to the basic idea of homogeneity and isotropy of the big bang 
universe
the {\it primordial} deuterium abundance should not vary
in space. One can only expect that the D abundance
decreases with cosmic time 
due to conversion of D into $^3$He and heavier elements in stars.
To check whether big bang nucleosynthesis [BBN] has occurred
homogeneously or not,
precise measurements
of absolute values of D/H at high redshift are extremely
important.
In a series of papers [3-6] we have shown that
this task is badly model dependent.
Here we present two examples to underline the
difficulties of the
inverse problem in the analysis of the H+D absorption blends.

\section{Results}

We consider a cloud of a thickness $L$ with a stochastic velocity
field but of homogeneous 
(H~I) density and
temperature. The velocity field is characterized 
by its rms amplitude  $\sigma_{t}$
and its correlation length $l > 0$.
The model is identical to that of [5].
This approach generalizes the standard procedure which assumes
no spatial correlations in the velocity field ($l \equiv 0$). -- 
To estimate physical parameters and an appropriate velocity field
structure along the line of sight, we used a Reverse Monte Carlo [RMC]
technique. The algorithm requires to define 
a simulation box for 5 physical parameters~:
N(H~I), D/H, $T_{kin}$, 
$\sigma_{t}/v_{th}$, and $L/l$
(here $v_{th}$ denotes the thermal width of the hydrogen lines).
 -- The continuous random function of 
the coordinate $v(s)$ is represented by
its sampled values at equal space intervals $\Delta s$, i.e. by
$\{v_1, v_2, \dots , v_k\}$, the
vector of the velocity components parallel to the line of sight
at the spatial points $s_j$ (for more detail, see [5]).

Direct observations of galactic halos at $z > 2$ [7] show that 
$\sigma_{t} \simeq 40 \pm 15$ km s$^{-1}$,  
if $T_{kin} \simeq 10^4$ K.
Our RMC calculations yield for the $z_a = 2.504$ and $z_a = 0.701$
absorption systems $\sigma_t \simeq 26$ km s$^{-1}$ [5,6] which makes
the procedure to be adequate, whereas $\sigma_t \simeq 2-8$ km s$^{-1}$
found in [1,2] and $\sigma_t \simeq 13$ km s$^{-1}$ found in [8,9]
are evidently too low.

To illustrate our results, we show in Fig.~1 H+D Ly$\alpha$ lines
observed by the two groups [1,2] and [8,9] and some of the adequate profile
fits ($\chi^2_{min}$ per degree of freedom $\simeq 1$). The solutions are
not unique, however, but depend sensitively on the velocity field
configuration along the line of sight (see [5,6], for more examples).
Fig.~1 shows that both profiles are compatible with D/H $\simeq
4.1\times10^{-5}$ (the uncertainty range is from $4.1\times10^{-5}$
to $4.6\times10^{-5}$ [5,6]). Taking the mean D/H value from this range,
one can estimate the baryon-to-photon ratio 
$\eta$ during SBBN and using the
present-day photon density determined from the cosmic microwave background
measurements, one can obtain the current baryon density $\Omega_b$~:  
$\eta \simeq 4.4\times10^{-10}$ and $\Omega_b h^2 \simeq 0.015$,
correspondingly.

We conclude that the reliability of the interpretation of deuterium
absorption line observations at high redshift is determined by two
factors~: ($i$) improvements in the detection equipment, and
($ii$) advances in the theory of line formation in turbulent media. 
\vspace{1pc}

{\it Acknowledgment} -- The authors are grateful to John Webb for
making available the calibrated HST/GHRS spectra of Q~1718+4807
and acknowledge helpful correspondence and comments by him and
Alfred Vidal-Madjar. This work was supported in part 
by the RFBR grant No. 96-02-16905a.

\section{References}

\vspace{1pc}

\re
1. Burles, S., Tytler, D., 1996, astro-ph 9603069.
\re
2. Burles, S., Tytler, D., 1997, {\it Astrophys.\ J.} (submit.),
astro-ph 9712109.
\re
3. Levshakov, S. A., Kegel, W. H., 1997, 
{\it Monthly\ Notices\ Roy.\ Astron.\ Soc.} {\bf 288}, 787.
\re
4. Levshakov, S. A., Kegel, W. H., Mazets, I. E., 1997, 
{\it Monthly\ Notices\ Roy.\ Astron.\ Soc.} {\bf 288}, 802.
\re
5. Levshakov, S. A., Kegel, W. H., Takahara, F., 1997, 
{\it Monthly\ Notices\ Roy.\ Astron.\ Soc.} 
(submit.), astro-ph 9710122.
\re
6. Levshakov, S. A., Kegel, W. H., Takahara, F., 1997, 
{\it Astron.\ Astrophys.} (submit.) .
\re
7. van Ojik, R., et.al., 1997, {\it Astron.\ Astrophys.}
{\bf 317}, 358.
\re
8. Webb, J. K., et.al., 1997, {\it Nature} {\bf 388}, 250.
\re
9. Webb, J. K., et.al., 1997, astro-ph 9710089.

\end{document}